\documentclass[10pt,twocolumn,oneside]{IEEEtran}
\usepackage{amsfonts}
\usepackage{amsfonts}
\usepackage{amsfonts}
\usepackage{amsfonts}
\usepackage{amsfonts}
\usepackage{amsfonts}
\usepackage{amsfonts}
\usepackage{amsfonts}
\usepackage{amsfonts}
\usepackage{amsfonts}
\usepackage{amsfonts}
\usepackage{amsfonts}
\usepackage{amsfonts}
\usepackage{bbm}
\usepackage{amsfonts}
\usepackage{amssymb}
\usepackage{stfloats}
\usepackage{cite}
\usepackage{graphicx}
\usepackage{psfrag}
\usepackage{subfigure}
\usepackage{amsmath}
\usepackage{array}
\usepackage{multirow}
\usepackage{tabularx}
\usepackage{color}
\usepackage{algorithm}

\begin{document}

\title{A Scalable Limited Feedback Design for Network MIMO using Per-Cell Product Codebook}

\newtheorem{Thm}{Theorem}
\newtheorem{Lem}{Lemma}
\newtheorem{Cor}{Corollary}
\newtheorem{Def}{Definition}
\newtheorem{Exam}{Example}
\newtheorem{Alg}{Algorithm}
\newtheorem{Prob}{Problem}
\newtheorem{Rem}{Remark}
\newtheorem{Proof}{Proof}
\newtheorem{Obs}{Observation}
\newtheorem{Con}{Conclusion}

\newcommand{\argmin}[1]{\underset{#1}{\operatorname{argmin}}\;}
\newcommand{\argmax}[1]{\underset{#1}{\operatorname{argmax}}\;}

\author{\IEEEauthorblockN{Yong Cheng, {\em Student Member, IEEE}, Vincent K. N. Lau, {\em Senior Member, IEEE}, and Yi Long}
\thanks{Manuscript received August 8, 2009; revised March 23, 2010
; accepted July 28, 2010. The editor coordinating the review of
this paper and approving it for publication was Prof. Davide Dardari.}
\thanks{Yong Cheng and Vincent K. N. Lau are with
the Department of Electronic and Computer Engineering (ECE), The
Hong Kong University of Science and Technology (HKUST), Hong Kong.
(email: \{chengy, eeknlau\}@ust.hk). }
\thanks{Yi Long is with Huawei Technology Co. Ltd., Shenzhen, China. (email: longy@huawei.com).}}

\maketitle

\begin{abstract}
In network MIMO systems, channel state information is required at
the transmitter side to multiplex users in the spatial domain. Since
perfect channel knowledge is difficult to obtain in practice,
\emph{limited feedback} is a widely accepted solution.
The {\em dynamic number of cooperating BSs} and {\em heterogeneous
path loss effects} of network MIMO systems pose new challenges on limited
feedback design. In this paper, we propose a scalable  limited feedback design for
network MIMO systems with multiple base stations, multiple users and
multiple data streams for each user. We propose a {\em limited feedback framework using per-cell product codebooks}, along with a {\em
low-complexity feedback indices selection algorithm}. We show that the proposed
per-cell product codebook limited feedback design can asymptotically achieve the
same performance as the joint-cell codebook approach. We also derive an asymptotic \emph{per-user throughput loss} due to limited feedback with
per-cell product codebooks. Based on that, we show that when the number of per-user feedback-bits $B_{k}$ is $\mathcal{O}\big( Nn_{T}n_{R}\log_{2}(\rho g_{k}^{sum})\big)$, the system operates in the \emph{noise-limited} regime in which the per-user throughput is $\mathcal{O} \left( n_{R} \log_{2} \big( \frac{n_{R}\rho g_{k}^{sum}}{Nn_{T}}  \big) \right)$. On the other hand, when the number of per-user feedback-bits $B_{k}$ does not scale with the \emph{system SNR} $\rho$, the system operates in the \emph{interference-limited} regime where the per-user throughput is $\mathcal{O}\left( \frac{n_{R}B_{k}}{(Nn_{T})^{2}} \right)$. Numerical
results show that the proposed design is very flexible to accommodate dynamic number of cooperating BSs and achieves much better performance compared with other baselines (such as the Givens rotation approach).
\end{abstract}


%\IEEEpeerreviewmaketitle

\section{Introduction}\label{sec:intro}
Network MIMO (multiple-input multiple-output) is considered as a
core technology for the next generation wireless systems. The key
idea of network MIMO is to employ base station (BS) cooperation
among the neighboring cells for joint signal transmission in
downlink direction and/or joint signal detection in uplink direction
\cite{Karakayali2006, Foschini2006, Andrews2007, Jing2008}. In
network MIMO systems, the undesirable inter-cell interference (ICI)
can be transformed into useful signals via collaborative
transmission among multiple adjacent BSs. Therefore, the network
MIMO solution could effectively leverage the advantage of MIMO
communications.

Similar to single-cell multiuser MIMO (MU-MIMO) communications,
knowledge of channel state information (CSI) is critical for
efficient spatial multiplexing of mobiles in network MIMO systems. Space
division multiple access (SDMA) for single-cell MU-MIMO has been
studied in lots of literatures \cite{Choi2004, Spencer2004,
WANG2007, Jindal2005, Yoo2007, Ravindran2008, Huang2009}. In
\cite{Choi2004, Spencer2004}, perfect knowledge of CSI at the
transmitter is assumed to eliminate cochannel interference
(CCI) among the users engaged in SDMA. However, perfect CSI is
difficult to obtain at the transmitter side in practice and there
are lots of literatures discussing SDMA with limited CSI feedback in
single-cell MU-MIMO systems \cite{Jindal2005, WANG2007, Yoo2007, Trivellato2007, Trivellato2008,
Ravindran2008, Huang2009}. Recently, the authors of \cite{Kim2008,
TSWJ08cmbo} have extended the single-cell limited feedback designs
to network MIMO systems by treating the cooperating BSs as a
composite MIMO transmitter (i.e., one super BS),  and this refers to the {\em joint-cell codebook} approach. While the existing
work \cite{Kim2008, TSWJ08cmbo} provide some preliminary solutions
for CSI feedback in network MIMO systems, there are still a number
of important issues to be addressed.
\begin{itemize}
\item{\textbf{Dynamic Number of Cooperating BSs:}}
One important difference between single-cell MIMO and network MIMO
systems is that the number of cooperating BSs in the latter case is
dynamic, depending on location of the mobiles. As a result, the total
number of bits for CSI feedback\footnote{In \textcolor{black}{LTE-Advanced} systems, the total
number of bits for limited feedback scales linearly with the number
of active BSs.} as well as the dimension of the CSI matrix seen by a
user are dynamic. The conventional limited feedback designs are all designed for fixed number of transmit antennas and cannot be directly applied to
network MIMO systems due to the lack of flexibility. In other words, it
is very important to have flexibility incorporated in the
codebook-based limited feedback schemes in network MIMO systems, so
that the same codebook can be used to quantize the CSI matrix seen
by a user regardless of the number of cooperating BSs. This poses a
new design criteria for limited feedback mechanisms in network MIMO
systems.
\item{\textbf{Heterogeneous Path Loss Effects\footnote{\textcolor{black}{"Heterogeneous path loss effects" refers to the different path losses from the $N$ cooperating BSs to one MS.}}:}}
In network MIMO systems, it is quite common to have non-uniform path
losses between a mobile station (MS) and the cooperating BSs.
Hence, the conventional Grassmannian codebooks \cite{Love2005,
Mondal2008, Dai2008}, which is designed to match the CSI matrix with i.i.d. entries, fail to
match the actual statistics of the \emph{aggregate CSI matrix}
seen by a user due to the heterogeneous path loss effects. In
addition, the path losses geometry seen by one MS are dynamic and
cannot be incorporated into the offline codebook design procedures.
\item{\textbf{Performance Analysis:}}
The analytical results of the single cell SDMA scheme with limited
feedback has been considered extensively in the literatures~\cite{Jindal2005, Yoo2007, Ravindran2008, Huang2009}. However,
these results cannot be applied to the multi-cell scenario with
limited feedback capturing the dynamic number of cooperating BSs and
the heterogeneous path loss effects.
\end{itemize}

One conventional limited feedback design, namely the Givens
rotation~\cite{Roh2004, Sadrabadi2006, Long2008}, could potentially
address the above challenges. Using Givens rotation, a unitary matrix (channel direction) is
decomposed into products of Givens matrices. Each Givens matrix
contains two Givens parameters, which are quantized using {\em
scalar} quantizer and fed back to the BSs. As a result, it offers
the flexibility because when the number of cooperating BSs changes,
the number of Givens matrices also changes accordingly. Hence, the
same scalar quantizer can be used to quantize unitary matrices of
time-varying dimensions. However, the issue of Givens rotation
approach is the poor feedback efficiency due to scalar (or two-dimensional vector) quantization.
In this paper, we shall propose a novel scalable limited  feedback mechanism using
\emph{per-cell product codebooks}\footnote{\textcolor{black}{"Per-cell product  codebook" refers to the codebook that is designed with the single BS antenna configuration; while "joint-cell codebook" refers to the codebook that is designed with the aggregate antenna configuration of the $N$ cooperating BSs. In other words, in joint-cell codebook design, the $N$ cooperating BSs are treated as one aggregate BS, which is called super-BS in the paper.}}  to address the dynamic
MIMO configurations and heterogeneous path loss effects, along with
a low-complexity realtime feedback indices selection algorithm. \textcolor{black}{
%In the proposed per-cell feedback framework, the traditional per-cell product codebooks that are designed %with single BS antenna configuration can be used in the network MIMO systems.
In the proposed feedback scheme, the \emph{product codebook} (defined in Section \ref{sec:formulation}) that is used for CSI feedback in the network MIMO systems is simply the \emph{Cartesian product} of $N$ per-cell product codebooks, with $N$ denoting the number of cooperating BSs. \emph{Cartesian product} operation allows for a single per-cell product codebook to be used irrespective of the number of cooperating BSs.} We shall show that the
proposed per-cell product codebook based limited feedback mechanism can asymptotically
achieve the same performance as the joint-cell codebook approach. We derive an asymptotic \emph{per-user throughput loss}
 due to limited feedback with
per-cell product codebooks. Based on the results, we show that when the number of per-user feedback-bits $B_{k}$ is $\mathcal{O}\big( Nn_{T}n_{R}\log_{2}(\rho g_{k}^{sum})\big)$, the system operates in a noise-limited regime with per-user throughput scaling as $\mathcal{O} \left( n_{R} \log_{2} \big( \frac{n_{R}\rho g_{k}^{sum}}{Nn_{T}}  \big) \right)$. On the other hand, when the number of per-user feedback-bits $B_{k}$ does not scale with the {\em system SNR} $\rho$, the system operates in an interference limited regime with per user throughput scaling as $\mathcal{O}\left( \frac{n_{R}B_{k}}{(Nn_{T})^{2}} \right)$. Numerical
results show while the proposed scheme is flexible to accommodate dynamic number of cooperating BS, it has significant performance gains over the reference baselines (e.g. Givens rotation approach).

The rest of this paper is organized as follows. We introduce the
network MIMO system model and codebook-based CSI feedback model in
Section~\ref{sec:sys_mod}. The proposed per-cell product codebook based
limited feedback framework is introduced in
Section~\ref{sec:Percell_LF}. Asymptotic performance analysis of the
proposed scheme is elaborated in Section~\ref{sec:perf_analy}. We
present the numerical results along with discussions in
Section~\ref{sec:num_res_disc}. Finally, we summarize the main results in Section~\ref{sec:conclusion}.

{\em Notations}: Matrices and vectors are denoted with boldface uppercase and lowercase letters, respectively; $\mathbf{A}^{\dag}$ and
$\mbox{tr}\left\{\mathbf{A}\right\}$   denote
the conjugate transpose and trace of matrix
$\mathbf{A}$, respectively; $\mathbf{I}_{L}$ represents the $L \times
L$ identity matrix; $\mathbb{CN}(\mu, \sigma^{2})$ denotes the
circularly symmetric complex Gaussian distribution,  with mean $\mu$
and variance $\sigma^{2}$; $\mathbb{C}$ and
$\mathbb{R}_{+}$ denote the set of complex numbers and positive real
numbers, respectively.

\section{System Model}\label{sec:sys_mod}
\subsection{Network MIMO Channel Model} \label{sec:channel_mod}
\begin{figure}
\centering
\includegraphics[scale=0.52]{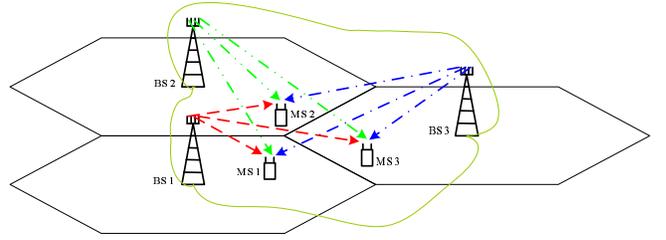}
\caption {An sample network MIMO system with $\left(n_{T},
 N, n_{R}, K\right)=\left(4, 3, 2, 6 \right)$. The 3 BSs
 collaboratively serve the 3 MSs via multi-cell SDMA} \label{fig:System_Model}
\end{figure}

Consider a network MIMO system $\left( n_{T}, N, n_{R}, K \right)$
with $N$ BSs serving $K$ active MSs in the downlink direction, as
shown in Fig. \ref{fig:System_Model}. The $N$ BSs are
inter-connected via high-speed backhauls, and collaboratively serve
the $K$ MSs through the standard SDMA scheme.
Without loss of
generality, we assume that each BS has $n_{T}$ antennas and each MS
has $n_{R}$ antennas\footnote{\textcolor{black}{Although we assume that the
cooperating BSs and active MSs have homogeneous antenna
configurations (i.e., every BS has $n_{T}$ transmit antennas, and every MS has $n_{R}$ receive antennas.), the proposed per-cell limited feedback mechanism can
be directly extended to the cases with heterogeneous antenna
configurations (i.e., different BSs and/or different MSs have different number of antennas.).}}. Assume {\em limited feedback based block
diagonalization} (LF-BD) \cite{WANG2007, Ravindran2008} is employed
for SDMA transmission in the network MIMO system, the downlink
signal model can be written as,
\begin{IEEEeqnarray}{rCl} \label{eqn:signal_model}
\mathbf{y}_{k} =
\mathbf{H}_{k}\mathbf{\widehat{W}}_{k}\sqrt{\mathbf{P}_{k}}\mathbf{d}_{k}
+\underbrace{\mathbf{H}_{k}\big(\sum_{j=1,j \neq
k}^{K}\mathbf{\widehat{W}}_{j}\sqrt{\mathbf{P}_{j}}\mathbf{d}_{j}\big)}_{\mbox{residual
CCI}} + \mathbf{z}_{k},
\end{IEEEeqnarray}
where $\mathbf{y}_{k}\in \mathbb{C}^{n_{R} \times 1}$ is the received signal
vector at the $k^{th}$ MS; $\mathbf{H}_{k}
\in\mathbb{C}^{n_{R}\times Nn_{T}}$ denotes the \emph{aggregate CSI matrix} of the $k^{th}$
MS; $\mathbf{\widehat{W}}_{k}\in \mathbb{C}^{Nn_{T} \times n_{R}}$
is the precoder for the $k^{th}$ MS, with
$\mathbf{\widehat{W}}_{k}^{\dag}\mathbf{\widehat{W}}_{k}=\mathbf{I}_{n_{R}}$;
$\mathbf{P}_{k}=\frac{NP_{max}}{Kn_{R}}\mathbf{I}_{n_{R}}$ is the power allocation
matrix for the $k^{th}$ MS, with $P_{max}$ representing the maximal transmit power of one BS; $\textbf{d}_{k}\in \mathbb{C}^{n_{R} \times 1 }$
is the transmitted symbol vector intended for the $k^{th}$ MS, satisfying
$\mathbb{E}\left\{ \mathbf{d}_{k} \mathbf{d}_{k}^{\dag}
\right\}=\mathbf{I}_{n_{R}}$; $\mathbf{z}_{k}\in \mathbb{C}^{n_{R} \times 1}$
denotes the noise vector, with i.i.d. $\mathbb{CN}(0, \sigma^{2})$
entries, i.e., $\mathbb{E}\left\{ \mathbf{z}_{k}
\mathbf{z}_{k}^{\dag} \right\}=\sigma^{2}\mathbf{I}_{n_{R}}$. In this paper, it is assumed that \textcolor{black}{$Kn_{R} \leq Nn_{T}$}.

For precoder design, LF-BD imposes the following
conditions on $\mathbf{\widehat{W}}_{k}$ to eliminate CCI at the
transmitter side \cite{WANG2007, Ravindran2008},
\begin{IEEEeqnarray}{rCl} \label{eqn:bd}
\mathbf{\widehat{H}}_{k}\mathbf{\widehat{W}}_{j}=\mathbf{0}_{n_{R}
\times L},~\forall~ k \neq j;\: k,j=1,2,\cdots, K,
\end{IEEEeqnarray}
where $\mathbf{\widehat{H}}_{k}$ is the quantized aggregated CSI matrix of the $k^{th}$ MS (see equation \eqref{eqn:Hkhat}).

In the network MIMO system, the \emph{aggregate CSI matrix} $\mathbf{H}_{k}$ seen by the $k^{th}$ MS can be partitioned into $N$ submatrices, i.e.,
\begin{IEEEeqnarray}{rCl} \label{eqn:Hk}
\mathbf{H}_{k}=\left[ \mathbf{H}_{k,1} \:\ \mathbf{H}_{k,2} \:\
\cdots \:\ \mathbf{H}_{k,N} \right], \forall~k = 1, 2, \cdots, K,
\end{IEEEeqnarray}
where $\mathbf{H}_{k,n} \in \mathbb{C}^{n_{R}\times n_{T}}$ denotes
the CSI matrix between the $n^{th}$ BS and the $k^{th}$ MS, which is
commonly modeled as \cite{Hongyuan2004},
\begin{IEEEeqnarray}{rCl} \label{eqn: H_model}
\mathbf{H}_{k,n} = \sqrt{g_{k,n}s_{k,n}}\mathbf{H}_{k,n}^{(w)},
\forall k=1,2,\cdots,K$; $n=1,2,\cdots,N, \notag \\
\end{IEEEeqnarray}
where $g_{k,n}\in \mathbb{R}_{+}$ denotes the path loss from the
$n^{th}$ BS to $k^{th}$ MS; $s_{k,n}\in \mathbb{R}_{+}$ denotes the
lognormal shadowing component; $\mathbf{H}_{k,n}^{(w)}
\in\mathbb{C}^{n_{R}\times n_{T}}$ denotes a random matrix with
i.i.d. $\mathbb{CN}(0,1)$ entries, i.e., the Rayleigh fading
component.

Moreover, the \emph{aggregate CSI matrix} $\mathbf{H}_{k}$ of the
$k^{th}$ MS can be factorized as,
\begin{IEEEeqnarray}{rCl} \label{eqn: ACM}
\mathbf{H}_{k}=
\mathbf{H}_{k}^{(w)}\mathbf{G}_{k}, ~\forall~k = 1, 2, \cdots, K,
\end{IEEEeqnarray}
where $\mathbf{H}_{k}^{(w)}\in\mathbb{C}^{n_{R}\times Nn_{T}}$
denotes a random matrix with i.i.d. $\mathbb{CN}(0,1)$ entries;
$\mathbf{G}_{k}\in\mathbb{R}^{Nn_{T}\times Nn_{T}}_{+}$ represents
the large-scale fading component, which is given by,
\begin{IEEEeqnarray}{rCl} \label{eqn:G}
\mathbf{G}_{k} = diag\bigg(
\big[\sqrt{g_{k,1}s_{k,1}}\mathbf{1}_{n_{T}} \:\
\sqrt{g_{k,2}s_{k,2}}\mathbf{1}_{n_{T}} \:\ \cdots \: \notag \\
\sqrt{g_{k,N}s_{k,N}}\mathbf{1}_{n_{T}}    \big] \bigg),
\end{IEEEeqnarray}
where $\mathbf{1}_{n_{T}}$ equals to $ \left[1 \: 1\: \cdots \: 1
\right]\in \mathbb{R}_{+}^{1 \times n_{T}}$, and $diag(\mathbf{a})$
denotes a diagonal matrix with diagonal entries given by the
elements of vector $\mathbf{a}$.

\subsection{Codebook-based Limited Feedback Model} \label{subsec:percell_CSI}
For downlink transmission with the LF-BD scheme, the essential
information (i.e., users channel subspaces) required at the BSs side
can be obtained through the \emph{codebook-based CSI feedback}
scheme. In a general codebook-based CSI feedback framework, the
design metric\footnote{For limited feedback based SDMA transmission,
a widely adopted design metric for CSI feedback schemes is
\emph{minimizing the inter-user interference}, rather than
maximizing the desired signal power \cite{WANG2007, Ravindran2008}.}
for minimizing the residual CCI is \emph{chordal
distance}\cite{WANG2007, Ravindran2008}, which is defined as,
\begin{IEEEeqnarray}{rCl} \label{eqn: chordal_dist}
d_{c}\left(\mathbf{V}_{1},~\mathbf{V}_{2}\right)=\frac{1}{\sqrt{2}}\left\|
\mathbf{V}_{1}\mathbf{V}_{1}^{\dag} -
\mathbf{V}_{2}\mathbf{V}_{2}^{\dag}\right\|_{F},
\end{IEEEeqnarray}
where $\mathbf{V}_{1},~\mathbf{V}_{2} \in \mathbb{C}^{Nn_{T} \times
n_{R}}$ are two orthonormal matrices\footnote{In this paper, an orthonormal matrix refers to a matrix with \emph{orthonormal columns}.}; and $\|\mathbf{A}\|_{F}$ denotes the
Frobenius norm of matrix $\mathbf{A}$.

Without loss of generality, we assume the BS and MS share the same CSI quantization codebook, denoted by $\varphi$. The $J^{th}$ codeword in the codebook $\varphi$ is given by $\mathbf{V}_{J}$ and the cardinality of the codebook $\varphi$  is $2^{B}$, with $B$ representing the number of feedback-bits. In the conventional single-cell MIMO communications, the codeword index $J^{*}$ reported by the MS is generated through the minimization of the \emph{chordal distance} between the quantization source $\mathbf{V}$ and the codedwords in the entire codebook $\varphi$. Mathematically,
\begin{IEEEeqnarray}{rCl} \label{eqn:convention_fb}
J^{*} = \argmin{\mathbf{V}_{J}\in \varphi }{}
d_{c}\left(\mathbf{V}_{J}, \mathbf{V}\right).
\end{IEEEeqnarray}

The following assumptions are made through the rest of this paper.
Firstly, all the MSs have perfect channel state information between the cooperative BSs and themselves. Secondly, the large-scale
fading component $\mathbf{G}_{k}$ is known at all the BSs and MSs within the network. Thirdly, each MS shall feed back the CSI knowledge to the cooperative BSs through zero-delay error-free feedback links. In addition, flat block-fading MIMO channel is assumed.

\section{Scalable Limited Feedback Mechanism based on Per-Cell Product Codebook} \label{sec:Percell_LF}
In the joint-cell codebook based CSI feedback schemes \cite{Kim2008,
TSWJ08cmbo}, the cooperating BSs are treated as one composite BS (super BS), and
the codebooks are designed for the \emph{aggregate CSI matrix} $\mathbf{H}_{k}$ in a standard way. \textcolor{black}{Moreover, authors of \cite{TSWJ08cmbo} has also considered the heterogeneous path loss effects and the effects of increasing cluster size with a variable number of cooperating BSs.} While the joint-cell codebook approaches provide some preliminary solutions for network MIMO
systems, two important issues, namely the dynamic number of cooperating BSs and the heterogeneous path loss effects, are ignored. In this section, we shall introduce a \emph{scalable limited feedback design based on per-cell product codebook} to accommodate the dynamic MIMO configurations
and deal with the heterogeneous path loss effects. Moreover, we shall formulate the realtime feedback indices determination (at the mobile) as a {\em combinatorial search} problem,
and propose a low-complexity solution.

\subsection{Scalable Per-cell Product Codebook based Limited Feedback Design} \label{sec:feedback_design}
In order to accommodate the dynamics of cooperating BSs and deal with the heterogeneous path
loss effects, we propose to quantize
$span\left\{\mathbf{H}_{k}^{(w)}\right\}$ with per-cell product codebooks,
rather than quantizing $span\left\{\mathbf{H}_{k}\right\}$ directly, where $span\left\{\mathbf{A}\right\}$ denotes the \emph{row space} of matrix $\mathbf{A}$. The proposed per-cell product codebook based CSI quantization procedure
involves both MS side processing and BS side processing, i.e.,
codewords indices generation at the MS side and quantized aggregate
CSI matrix reconstruction at the BS side. Denote the $N$ per-cell product
codebooks as
$\varphi_{1},\varphi_{2},\cdots,\varphi_{N}$,  where $\varphi_{n} = \left\{\mathbf{V} \big| \mathbf{V} \in \mathbb{C}^{n_{T} \times n_{R}}, \mathbf{V}^{\dag}\mathbf{V} = \mathbf{I}_{n_{R}}\right\}$, with cardinality $2^{B_{k,n}}$, and the total number of per-user feedback-bits $B_{k}$ is given by $B_{k}= \sum_{n = 1}^{N}B_{k,n}$. Note that the $N$ per-cell product  codebooks could be identical. The proposed scalable limited feedback processing at the MS and the BS sides is summarized below:

\emph{Feedback Indices  Generation (MS side)}:  take the $k^{th}$ MS as an
example. The feedback indices generation is a mapping from the normalized aggregated CSI matrix $\textbf{H}_{k}^{(w)}$ to the feedback indices $\big\{ J_{k,1}^{*},$ $ J_{k,2}^{*}, \:
\cdots, \: J_{k,N}^{*} \big\}$ corresponding to the $N$ per-cell product codebooks $\varphi_{1}, \varphi_{2},\cdots, \varphi_{N}$.
\begin{itemize}
\item{\textbf{Normalization}:} To handle the heterogeneous path loss effects,
we first normalize the aggregate CSI matrix $\mathbf{H}_{k}$ by
the large-scale fading component $\mathbf{G}_{k}$, i.e.,
\begin{IEEEeqnarray}{rCl} \label{eqn:normalization}
\mathbf{H}_{k}^{(w)} = \mathbf{H}_{k}\mathbf{G}_{k}^{-1}.
\end{IEEEeqnarray}

\item{\textbf{Decomposition}:} Apply the standard \emph{singular value decomposition} (SVD) \cite{Tse05} to the normalized CSI matrix $\mathbf{H}_{k}^{(w)}$, we have,
\begin{IEEEeqnarray}{rCl} \label{eqn:svd_Hw}
 \mathbf{H}_{k}^{(w)} &=&
\mathbf{U}_{k}^{(w)} \left[ \mathbf{S}_{k}^{(w)} \
\mathbf{0}_{n_{R}\times {Nn_{T}-n_{R}} } \right]
\left[\mathbf{V}_{k}^{(w)} \  \mathbf{\tilde{V}}_{k}^{(w)}
\right]^{\dag} \notag \\ &=& \mathbf{U}_{k}^{(w)} \mathbf{S}_{k}^{(w)}
\left[\mathbf{V}_{k}^{(w)} \right]^{\dag}.
\end{IEEEeqnarray}

\item{\textbf{Realtime Feedback Indices Selection:}}
$\mathbf{V}_{k}^{(w)}$ is then mapped into $N$ per-cell product
codebooks with indices $\big\{ J_{k,1}^{*},$ $ J_{k,2}^{*}, \:
\cdots, \: J_{k,N}^{*} \big\}$, and these codewords indices are fed back to the BSs via an error-free feedback link.  The realtime indices generation is formulated as a combinatorial optimization in Section \ref{sec:formulation}.
\end{itemize}

\emph{Aggregate CSI Matrix Reconstruction and SDMA Precoder Computation (BS side)}: After collecting all the indices from the $K$ MSs, the BSs try to reconstruct the CSI matrices and compute the SDMA precoder from the $N$ codeword indices $\big\{ J_{k,1}^{*},$ $ J_{k,2}^{*}, \:
\cdots, \: J_{k,N}^{*} \big\}$.
\begin{itemize}
\item{\textbf{Reconstruction}:}
The indices $\left\{J_{k,1}^{*}, J_{k,2}^{*}, \cdots,
J_{k,N}^{*}\right\}$  from the $k^{th}$ MS are used
to construct a quantized version of $\mathbf{V}_{k}^{(w)}$ (denoted
as $\mathbf{\widehat{V}}_{k}^{(w)}$), i.e.,
\begin{IEEEeqnarray}{rCl}
\mathbf{\widehat{V}}_{k}^{(w)}=\frac{1}{\sqrt{N}}\left[\mathbf{V}_{J_{k,1}^{*}}^{\dag}
\; \mathbf{V}_{J_{k,2}^{*}}^{\dag} \; \cdots \;
\mathbf{V}_{J_{k,N}^{*}}^{\dag} \right]^{\dag}.
\end{IEEEeqnarray}

\item{\textbf{Denormalization}:}
$\mathbf{\widehat{V}}_{k}^{(w)}$ is used to construct the
quantized aggregate CSI matrix $\mathbf{\widehat{H}}_{k}$ of the $k^{th}$ MS, which
is given by,
\begin{IEEEeqnarray}{rCl} \label{eqn:Hkhat}
\mathbf{\widehat{H}}_{k}=\left[\mathbf{\widehat{V}}_{k}^{(w)}\right]^{\dag}\mathbf{G}_{k},
~ \forall~k=1,2,\cdots, K.
\end{IEEEeqnarray}
\item{\textbf{SDMA Precoder Computation}:} Apply the standard SVD to the quantized interference channel $\mathbf{\widehat{H}}_{-k}$ seen by the $k^{th}$ MS, we have,
\begin{IEEEeqnarray}{rCl} \label{eqn:SVDHk}
\mathbf{\widehat{H}}_{-k} = \mathbf{\widehat{U}}_{-k}\mathbf{\widehat{S}}_{-k} \left[ \mathbf{\widehat{V}}_{-k}  \:\ \mathbf{\widetilde{\widehat{V}}}_{-k}\right]^{\dag},
\end{IEEEeqnarray}
where $\mathbf{\widehat{H}}_{-k}$ is given by,
\begin{IEEEeqnarray}{rCl} \label{eqn:SDMA_Precoder}
\mathbf{\hat{H}}_{-k}=\left[ \mathbf{\hat{H}^{\dag}}_{1} \:\
\mathbf{\hat{H}^{\dag}}_{2} \: \cdots \:
\mathbf{\hat{H}^{\dag}}_{k-1} \:\ \mathbf{\hat{H}^{\dag}}_{k+1}
\: \cdots \:
 \mathbf{\hat{H}^{\dag}}_{K} \right]^{\dag}.
\end{IEEEeqnarray}

The SVD operation \eqref{eqn:SVDHk} generates an orthonormal basis of the \emph{right null space} of $\mathbf{\widehat{H}}_{-k}$, i.e., $\mathbf{\widetilde{\widehat{V}}}_{-k} \in \mathbb{C}^{Nn_{T} \times n_{R}}$.  We can set $\mathbf{\widehat{W}}_{k} = \mathbf{\widetilde{\widehat{V}}}_{-k}$, which satisfies equation \eqref{eqn:bd}.
\end{itemize}

\begin{Rem} [Ways of Sending back Feedback Indices]
\textcolor{black}{When the $k^{th}$ MS sends back the feedback indices $\big\{ J_{k,1}^{*},$ $ J_{k,2}^{*}, \:
\cdots, \: J_{k,N}^{*} \big\}$, it could the feedback index $J_{k,n}^{*}$ the $n^{th}$ BS, $\forall n = 1, 2, \cdots, N$; or it could send all the feedback indices $\big\{ J_{k,1}^{*},$ $ J_{k,2}^{*}, \:
\cdots, \: J_{k,N}^{*} \big\}$ to its nearest BS; or it could simply broadcast all the feedback indices $\big\{ J_{k,1}^{*},$ $ J_{k,2}^{*}, \:
\cdots, \: J_{k,N}^{*} \big\}$, which will be received by all the $N$ cooperating BSs thanks to the broadcast nature of the wireless media.}
\end{Rem}

\begin{Rem} [Generalization of Common Cooperating BSs]
Without loss of generality, we have assumed that the $K$ MSs have the same set of cooperating BSs. The above per-cell product codebook limited feedback framework can also be applied directly to the case where each MS has a different active cooperating BSs set. For example, suppose MS-1 has BS-1 and BS-2 as its active set and MS-2 has BS-2 and BS-3 as its active set. This can be accommodated by our framework by considering a common active set of BS-1, BS-2, BS-3 for both MS-1 and MS-2 and setting $g_{1, 3} = g_{2, 1} = 0$.
\end{Rem}

As a summary, the  proposed per-cell product codebook based limited feedback mechanism has
the following advantages.
\begin{itemize}
\item The proposed scheme relies on per-cell product codebooks, which are designed offline based on single-BS MIMO configurations. The proposed scheme is scalable w.r.t. any number of cooperating BSs.
\item Standard precoder codebooks (such as the Grassmannian codebook, Lloyd's codebook, etc.) can be used in the proposed framework because the heterogeneous path loss issue is handled realtime in equation \eqref{eqn:normalization} and \eqref{eqn:Hkhat}.
\item We could further exploit the special structure of $N$ per-cell product codebooks in the proposed framework to derive a low complexity feedback indices selection algorithm.
\end{itemize}

\subsection{Problem Formulation} \label{sec:formulation}
The feedback indices generation at the MS side is non-trivial, since it involves {\em combinatorial search} over the $N$ per-cell product codebooks. In the following, we shall first define the {\em aggregate codeword} for the $N$ per-cell product codebooks and then formulate the {\em feedback indices generation problem} as a {\em chordal distance minimization problem} between the {aggregate-codeword} and the quantization source.
\begin{Def}[aggregate-codeword] \label{def:MW_SINR}
Let $\mathbf{V}_{J_{k,n}} \in \mathbb{C}^{n_{T} \times n_{R}}$
denote the $J_{k,n}\mbox{-th}$ codeword in codebook $\varphi_{n}$,
an \emph{aggregate-codeword} $\mathbf{\bar{V}}\left(J_{k,1},
J_{k,2},\cdots,J_{k,N}\right)$ is defined to be,
\begin{IEEEeqnarray}{rCl} \label{eqn:aggregate}
\mathbf{\bar{V}}\left(J_{k,1},
J_{k,2},\cdots,J_{k,N}\right)=\frac{1}{\sqrt{N}}\left[\mathbf{V}_{J_{k,1}}^{\dag}
\; \mathbf{V}_{J_{k,2}}^{\dag} \; \cdots \;
\mathbf{V}_{J_{k,N}}^{\dag} \right]^{\dag}, \notag \\
\end{IEEEeqnarray}
where $\mathbf{V}_{J_{k,n}} \in \varphi_{n}$, $\forall n = 1, 2, \cdots, N$. ~\hfill\IEEEQEDclosed
\end{Def}

By the definition of the aggregate-codeword, the feedback indices at $k^{th}$ MS side can be determined through the following optimization problem.
\begin{Prob} [Optimal Feedback Indices Generation] \label{prob:jcis}
Finding out $N$ codewords indices, that are denoted as $\big\{J_{k,1}^{*}, \: J_{k,2}^{*}, \:
\cdots, \: J_{k,N}^{*}\big\}$, in the $N$ per-cell product codebooks
$\varphi_{1},\varphi_{2},\cdots,\varphi_{N}$ respectively,  such
that the {\em chordal distance} between the
\emph{aggregate-codeword} $\mathbf{\bar{V}}\left(J_{k,1},
J_{k,2},\cdots,J_{k,N}\right)$ and the quantization source
$\mathbf{V}_{k}^{(w)}$ is minimized. Mathematically, the feedback indices generation problem can be modeled as the following optimization problem,
\begin{IEEEeqnarray}{rCl} \label{eqn:JCIS}
%&\left\{J_{k,1}^{*}, \: J_{k,2}^{*}, \: \cdots, \: J_{k,N}^{*}\right\} = \notag \\
&\min_{J_{k,1},  J_{k,2}, \:\cdots, J_{k,N}
}{}  &{d_{c}\left(\mathbf{\bar{V}}\left(J_{k,1},
J_{k,2},\cdots,J_{k,N}\right), \mathbf{V}_{k}^{(w)} \right)} \\
&\textrm{subject to} &\mathbf{V}_{J_{k,n}} \in \varphi_{n},  \forall n = 1,2,
\cdots, N. \label{eqn:constraint}
\end{IEEEeqnarray}
\end{Prob}

\textcolor{black}{In the above proposed per-cell product codebook based limited feedback scheme, the aggregate codeword $\mathbf{\bar{V}}\left(J_{k,1},
J_{k,2},\cdots,J_{k,N}\right)$ can be thought as a codeword in the \emph{product codebook} $\varphi_{Per} = \varphi_{1}\otimes  \varphi_{2} \otimes   \cdots \otimes \varphi_{n}$, i.e.,
\begin{IEEEeqnarray}{rCl} \label{eqn:R_Pro_Cbk}
\mathbf{\bar{V}}\left(J_{k,1},
J_{k,2},\cdots,J_{k,N}\right) \in \varphi = \varphi_{1}\otimes  \varphi_{2} \otimes   \cdots \otimes \varphi_{n},
\end{IEEEeqnarray}
where $\mathbf{V}_{J_{k,n}} \in \varphi_{n}$, $\forall n = 1, 2, \cdots, N$. The product codebook $\varphi$ is important because it allows for a single codebook to be designed, and the real codebook that is used for CSI feedback is simply the Cartesian product of $N$ single cell codebooks $\left\{\varphi_{n}\right\}_{n=1}^{N}$.}

\begin{Rem} [Backward Compatibility]
When $N$ equals $1$, i.e., the single-BS scenario,
\emph{Problem
\ref{prob:jcis}} will degenerate to the conventional feedback index generation problem.
%, as shown in \eqref{eqn:convention_fb}
\end{Rem}

\subsection{Low-complexity Solution}\label{sec:local_search}
\emph{Problem \ref{prob:jcis}} belongs to the standard combinatorial search problem \cite{Papadimitriou1998} and the optimal solution requires \emph{exhaustive search} over the
$N$ per-cell product codebooks, which has exponential complexity w.r.t. the number of feedback-bits $B_{k}$.
However, in the practical communication systems, a MS may not be
able to support such complicated operations. In order to address
this issue, we shall propose a low-complexity searching algorithm,
which exploit the per-cell product codebook structure and decomposes the searching process over the $N$ codebooks into
the searching over $N$ {\em sub-codebooks} with reduced size. For illustration
purpose, we shall first give the definition of {\em sub-codebook}.
\begin{Def} [Sub-codebook]  \label{subcbk_def}
A \emph{sub-codebook} $\bar{\varphi}\left(\mathbf{V}_{k,n}^{(w)}, \delta_{n}\right)$,
is defined as a collection of codewords in the original codebook $\varphi_{n}$, which lies in the neighborhood of $\delta_{n}$ of the quantization source $\mathbf{V}_{k,n}^{(w)}$. Mathematically, we have\footnote{In this paper, we use the {\em chordal distance} as the distance measure. Other distance metrics can also be applied here.},
\begin{IEEEeqnarray}{rCl} \label{eqn:metric_ball}
\bar{\varphi}\left(\mathbf{V}_{k,n}^{(w)}, \delta_{n}\right) \triangleq
\left\{\mathbf{V} \left| \mathbf{V} \in \varphi_{n};
d_c\left(\mathbf{V},\mathbf{V}_{k,n}^{(w)}\right) < \delta_{n}
\right. \right\}.
\end{IEEEeqnarray}
\end{Def}

Based on {\em Definition \ref{subcbk_def}}, we can propose our low-complexity searching algorithm as well as complexity analysis in the following (see {\em Algorithm \ref{alg: low_complex}} below).
\begin{algorithm}
\caption{\textcolor{black}{Low-complexity Indices Selection Algorithm (ISA)}} \label{alg: low_complex}
\begin{itemize}
\item \textcolor{black}{\textbf{Step 1: Find Localized Centroids}:} \\
\textcolor{black}{At the $k^{th} (\forall k = 1, 2, \cdots, K)$ MS, the standard SVD operation is first applied to the CSI matrix $\mathbf{H}_{k,n}^{(w)}$, i.e.,
\begin{IEEEeqnarray}{rCl} \label{eqn:Alg_SVD}
\mathbf{H}_{k,n}^{(w)} &=& \mathbf{U}_{k,n}^{(w)} \left[\mathbf{S}_{k,n}^{(w)} \:\ \mathbf{0}_{n_{R} \times (n_{T}-n_{R})} \right]  \left[\mathbf{V}_{k,n}^{(w)} \:\ \mathbf{\widetilde{V}}_{k,n}^{(w)}\right] \notag \\ &=& \mathbf{U}_{k,n}^{(w)} \mathbf{S}_{k,n}^{(w)} \mathbf{V}_{k,n}^{(w)}.
\end{IEEEeqnarray}
Then, the $n_{T} \times (n_{T}-n_{R})$ orthonormal basis $\mathbf{V}_{k,n}^{(w)}$ of the row space of the CSI matrix $\mathbf{H}_{k,n}^{(w)}$ is set as the centroid of the sub-codebook $\bar{\varphi}\left(\mathbf{V}_{k,n}^{(w)}, \delta_{n}\right)$. In this step, the standard SVD operation is with the time complexity of $\mathcal{O}\left({n_{T}n_{R}^{2}}\right)$.}
\item \textcolor{black}{\textbf{Step 2: Construct \emph{Sub-codebooks}}:} \\
\textcolor{black}{After finding out the centroid $\mathbf{V}_{k,n}^{(w)}$ of the sub-codebook $\bar{\varphi}\left(\mathbf{V}_{k,n}^{(w)}, \delta_{n}\right)$, the $k^{th} (\forall k = 1, 2, \cdots, K)$ MS proceed to select all the codewords in codebook $\varphi_{n}$ that are within the neighborhood $\delta_{n}$  of the centroid $\mathbf{V}_{k,n}^{(w)}$, to construct sub-codebook $\bar{\varphi}\left(\mathbf{V}_{k,n}^{(w)}, \delta_{n}\right)$. Specifically,
\begin{IEEEeqnarray}{rCl} \label{eqn:Alg_Sub_Cbk}
&&\mbox{If } \mathbf{V} \in \varphi_{n}  \mbox{ and } d_c\left(\mathbf{V},~ \mathbf{V}_{k,n}^{(w)}\right) < \delta_{n}, \notag \\
&&~~~\mbox{       then } \mathbf{V} \in \bar{\varphi}\left(\mathbf{V}_{k,n}^{(w)}, \delta_{n}\right), \forall n = 1, 2, \cdots, N,\notag
\end{IEEEeqnarray}
where the time complexity of this step  is of the order $\mathcal{O}\left(\sum_{n=1}^{N}2^{B_{k,n}}\right)$.
}
\item \textcolor{black}{\textbf{Step 3: Indices Selection with Sub-codebooks}:} \\
\textcolor{black}{The $k^{th} (\forall k = 1, 2, \cdots, K)$  MS then searches the feedback indices within the restricted sub-codebooks $\left\{\bar{\varphi}\left(\mathbf{V}_{k,n}^{(w)}, \delta_{n}\right)\right\}_{n=1}$ to solve {\em Problem \ref{prob:jcis}}. Specifically, the $k^{th}$ MS tries to find out the feedback indices $\left\{J_{k,1}^{*}, \: J_{k,2}^{*}, \: \cdots, \: J_{k,N}^{*}\right\}$ through solving the following optimization problem with exhaustive search:
\begin{IEEEeqnarray}{rCl} \label{eqn:Alg_JCIS}
&\min_{J_{k,1},  J_{k,2}, \cdots,  J_{k,N}}{}  &{d_{c}(\mathbf{\bar{V}}\left(J_{k,1},
J_{k,2},\cdots,J_{k,N}\right), \mathbf{V}_{k}^{(w)} )} \notag \\
&\textrm{subject to} &\mathbf{V}_{J_{k,n}} \in \bar{\varphi}\left(\mathbf{V}_{k,n}^{(w)}, \delta_{n}\right),
\end{IEEEeqnarray}
where the time complexity of this step is of the order of $\mathcal{O}\left(\prod_{n=1}^{N}\left|\bar{\varphi}\left(\mathbf{V}_{k,n}^{(w)}, \delta_{n}\right)\right|\right)$,with $\left|\bar{\varphi}\left(\mathbf{V}_{k,n}^{(w)}, \delta_{n}\right)\right|$ denoting the cardinality of the sub-codebook $\bar{\varphi}\left(\mathbf{V}_{k,n}^{(w)}, \delta_{n}\right)$, which depends on $\delta_{n}$ and $B_{k,n}$. }
\end{itemize}
\end{algorithm}

\begin{Rem} [Performance-Complexity Tradeoff of $\delta_{n}$] In the above algorithm, the value of $\delta_{n}$ can be utilized to
tradeoff the average quantization distortion performance and the computational
complexity. In particular, when $\delta_{1}=\delta_{2}=\cdots
=\delta_{N}=\sqrt{n_{R}}$, the above algorithm reduces to  the
traditional exhaustive search algorithm. \textcolor{black}{As long as $\delta_{n} \leq \sqrt{n_{R}}$, $\forall n = 1, 2, \cdots, N$, then $\prod_{n=1}^{N} \left|\bar{\varphi}\left(\mathbf{V}_{k,n}^{(w)}, \delta_{n}\right)\right| \leq 2^{ \sum_{n=1}^{B_{k,n}} } = 2^{ B_{k} }$, which is the time complexity of the exhaustive search method for solving {\em Problem \ref{prob:jcis} } with the original codebooks $\left\{\varphi_{n}\right\}_{n=1}^{N}$.}
\end{Rem}

\section{Asymptotic Performance Analysis}\label{sec:perf_analy}
In this section, we shall quantify the asymptotic performance of the proposed per-cell product codebook based limited feedback mechanism w.r.t. the system configurations $(n_{T}, N, n_{R}, K)$ and the per-user feedback-bits $B_{k}$. In order to have tractable analysis so as to obtain design insights, we shall analyze the performance of the proposed limited feedback design using \emph{random codebooks} \cite{Ravindran2008, Dai2008}.

\subsection{Asymptotic Optimality}
We start with comparing the asymptotic performance of the proposed per-cell product codebook based limited feedback scheme and the joint-cell codebook approach. Denote $\varphi_{Joint}$ and $\Phi_{Joint}$ as a random joint-cell codebook and the collection of all possible random joint-cell codebooks, respectively. Similarly, denote $\varphi_{Per}$ and $\Phi_{Per}$ as a random  \emph{product per-cell product codebook} and the collection of all possible random  \emph{product per-cell product codebooks}, respectively, where a random  \emph{product per-cell product codebook} is defined as the \emph{Cartesian product} of $N$ random per-cell product codebooks, i.e., $\varphi_{Per} = \varphi_{1} \otimes \varphi_{2} \otimes \cdots \otimes \varphi_{N}$.  Let $\bar{D}_{k}\left( \Phi_{Joint} \right)$ and $\bar{D}_{k}\left( \Phi_{Per} \right)$ denote the  \emph{average quantization
distortion} averaged over all possible random joint-cell codebooks and random per-cell product codebooks, respectively, which are defined as,
\begin{IEEEeqnarray}{rCl} \label{eqn: ave_dc}
&&\bar{D}_{k}\big( \Phi_{Joint} \big) \triangleq \notag \\
&&\mathbb{E}\left\{ \min_{\mathbf{V} \in \varphi_{Joint} } d_{c}^{2}\big( \mathbf{V} , ~\mathbf{V}_{k}^{(w)}
\big) \big| \mathbf{H}_{k}^{(w)} ; \varphi_{Joint} \in \Phi_{Joint} \right\} , \label{eqn:ave_dist_joint}\\
&&\bar{D}_{k}\big( \Phi_{Per} \big) \triangleq \notag \\
&&\mathbb{E}\left\{ \min_{\mathbf{V} \in \varphi_{Per} } d_{c}^{2}\big( \mathbf{V} , ~\mathbf{V}_{k}^{(w)}
\big) \big| \mathbf{H}_{k}^{(w)} ; \varphi_{Joint} \in \Phi_{Per}\right\}. \label{eqn:ave_dist_per}
\end{IEEEeqnarray}

In order to show the efficiency of the proposed per-cell product codebook based limited feedback scheme, we shall establish the \emph{asymptotic optimality}
of the proposed limited feedback design w.r.t. the
joint-cell codebook approach and summarize the main results in the following
Lemma.
\begin{Lem} [Asymptotic Optimality] \label{lem: add_const}
For sufficiently large $n_{T}$ and finite $N$, we have:
\begin{itemize}
\item[(I)] \textcolor{black}{ The $Nn_{T} \times n_{R}$ orthonormal basis $\mathbf{V}_{k}^{(w)}$  of the row-space of the $n_{R} \times Nn_{T}$ normalized CSI matrix $\mathbf{H}_{k}^{(w)}$ has the same structure as the aggregate-codeword defined in \eqref{eqn:aggregate} almost surely (i.e., with probability $1$);}
\item[(II)] \textcolor{black}{The proposed per-cell
product codebook based limited feedback scheme and the joint-cell codebook approach achieve the same  \emph{average
quantization distortion}, i.e.,
\begin{IEEEeqnarray}{rCl} \label{eqn:asymptotic_optimality}
\bar{D}_{k}\left( \Phi_{Per} \right) = \bar{D}_{k} \left( \Phi_{Joint} \right), ~\forall~ k = 1,2, \cdots, K.
\end{IEEEeqnarray}}
\end{itemize}
\end{Lem}
\begin{proof}
Please refer to Appendix \ref{app: L2_add_const} for the proof.
\end{proof}

By virtue of {\em Lemma \ref{lem: add_const}}, we can derive the \emph{average quantization
distortion} associated with the random per-cell product codebooks, which is summarized in the following lemma.
\begin{Lem} [Average Quantization Distortion] \label{lem:approx_error}
For sufficiently large $B_{k}$, $n_{T}$, and small $n_{R}$,  the average
quantization distortion associated with the random per-cell product codebooks is given by,
\begin{IEEEeqnarray}{rCl} \label{eqn:approx_error}
\bar{D}_{k}\left( \Phi_{per} \right )  \approx
n_{R}2^{-\frac{B_{k}}{n_{R}\left(Nn_{T}-n_{R}\right)}}.
\end{IEEEeqnarray}
\end{Lem}
\begin{proof}
Please refer to Appendix \ref{app:approx_error} for the proof.
\end{proof}

\begin{Rem}[Average Quantization Distortion] \textcolor{black}{In {\em Lemma \ref{lem:approx_error}} the average
quantization distortion is associated with quantizing the row-space of the normalized CSI matrix $\mathbf{H}_{k}^{(w)}$, which consists of $\mathbb{C}\mathbb{N}(0,1)$ entries. As a result, the expression given in \eqref{eqn:approx_error} is the same as equation (8) given in reference \cite{Ravindran2008}, which is in fact first proved in reference \cite{Dai2008}.}
\end{Rem}

In the rest of this section, we shall derive the asymptotic
performance of the proposed limited feedback design based on the above two lemmas, and study the effects of limited
feedback and the advantage of macrodiversity provided by BS cooperation.

\subsection{Effect of Limited Feedback}\label{subsec: throughput_ana}
Within the framework of limit feedback study, the \emph{throughput
loss} due to CSI quantization is a common performance measure
\cite{Jindal2005, Yoo2007, Ravindran2008, Huang2009}.  In this paper, we
extend the concept into network MIMO configuration and define {\em
per-user throughput loss} $R_{k}^{Loss}$ as follows.
\begin{Def} [Per-user Throughput Loss] The per-user throughput
loss $R_{k}^{Loss}$ (w.r.t. the $k^{th}$ MS) is defined as the
throughput gap between the global CSI\footnote{In this paper,
global CSI means that all the cooperating BSs has perfect
CSI knowledge of the whole network.} (GCSI) case and the proposed per-cell product codebook based limited
feedback design, i.e.,
\begin{IEEEeqnarray}{rCl} \label{eqn:R_Loss_Def}
R_{k}^{Loss} = R_{k}^{CSIT} - R_{k}^{LF}, ~\forall~k =1, 2, \cdots, K,
\end{IEEEeqnarray}
\end{Def}
where \textcolor{black}{CSIT is the abbreviation for Channel State Information at the Transmitter Side and LF is the abbreviation for Limited Feedback}. Moreover,
\begin{IEEEeqnarray}{rCl} \label{eqn:R_CSIT}
&&R_{k}^{CSIT} =
\mathbb{E}\left\{\log_{2}\det\left(\mathbf{I}_{n_{R}}+\frac{1}{\sigma^{2}}\mathbf{H}_{k}\mathbf{W}_{k}\mathbf{P}_{k}\mathbf{W}_{k}^{\dag}\mathbf{H}_{k}^{\dag}\right)\right\},
\\
&&R_{k}^{LF} =
\mathbb{E}\bigg\{\log_{2}\det\bigg(\mathbf{I}_{n_{R}}+\bigg(\sigma^{2}\mathbf{I}_{n_{R}}+ \notag \\
&&~~~\mathbf{H}_{k}\sum_{j=1,j\neq
k}^{K}\mathbf{\widehat{W}}_{j}\mathbf{P}_{j}\mathbf{\widehat{W}}_{j}^{\dag}\mathbf{H}_{k}^{\dag}\bigg)^{-1}\mathbf{H}_{k}\mathbf{\widehat{W}}_{k}\mathbf{P}_{k}\mathbf{\widehat{W}}_{k}^{\dag}\mathbf{H}_{k}^{\dag}\bigg)\bigg\}, \label{eqn:RkLF}
\end{IEEEeqnarray}
with $\mathbf{W}_{k}$ denoting the percoder for the $k^{th}$ MS designed based on GCSI. Note that the residual CCI term in equation \eqref{eqn:RkLF} is due to limited feedback effects, and the expectation operation is taken over Rayleigh fading and the lognormal shadowing effect, as well as the random per-cell product codebooks (for $R_{k}^{LF}$ only). ~\hfill\IEEEQEDclosed

The following theorem quantifies the asymptotic per-user
throughput loss of the proposed limited feedback scheme
w.r.t. the network configuration $(n_{T}, N, n_{R}, K)$ , the
per-user feedback-bits $B_{_k}$ and the path loss geometry $\big \{g_{k,1}, g_{k,2}, \cdots, g_{k,N} \big\}$, \textcolor{black}{which consists of all the path losses between the $N$ cooperating BSs and the $k^{th}$ MS, and $\left\{g_{k,n}\right\}$ has been normalized to the weakest path so that $g_{k,n}\geq 1$.}

\begin{Thm} [Asymptotic Per-user Throughput Loss] \label{thm:rate_loss}
In the network MIMO system with the proposed  \emph{per-cell product codebook based limited
feedback scheme}, the asymptotic per-user throughput loss is given by,
\begin{IEEEeqnarray}{rCl} \label{eqn:R_loss}
R_{k}^{Loss} = \mathcal{O} \left( n_{R}\log_{2} \big(  2^{-\frac{B_{k}}{n_{R}(Nn_{T}-n_{R})}} \rho g_{k}^{sum} \big) \right),
\end{IEEEeqnarray}
where $\rho = \frac{P_{max}}{\sigma^{2}}$ is termed as {\em system SNR}; $g_{k}^{sum}$ is defined to be $\sum_{n=1}^{N}g_{k,n}$.
\end{Thm}
\begin{proof}
Please refer to Appendix \ref{app: T1_rate_loss} for the proof.
\end{proof}

As direct consequences of {\em Theorem \ref{thm:rate_loss}}, we have the following corollaries.
\begin{Cor} [Scaling Law for the Noise-Limited Regime] \label{thm:Bk_scale}
For the per-cell product codebook based feedback scheme, the per-user
feedback-bits $B_{k}$ required to bound the per-user throughput loss within a constant $\varepsilon$ shall scale according to the following expression,
\begin{IEEEeqnarray}{rCl} \label{eqn:Bk_approx}
B_{k} \approx  n_{R}( Nn_{T}-n_{R} ) \log_{2}
\big(\rho g_{k}^{sum} \big) - c(\varepsilon),
\end{IEEEeqnarray}
where $c(\varepsilon)=n_{R}( Nn_{T}-n_{R} )\log_{2}\big( 2^{\frac{\varepsilon}{n_{R}}
}-1\big)$.
\end{Cor}
\begin{proof}
Setting \textcolor{black}{right hand side (RHS)} of equation \eqref{eqn:prf_RkLoss} in {\em Appendix \ref{app: T1_rate_loss}}  to $\varepsilon$, and solving for $B_{k}$ will result in equation \eqref{eqn:Bk_approx} directly.
\end{proof}

\begin{Cor} [Scaling Law for the Interference-Limited Regime] \label{thm:rate_const_IFL}
For the proposed per-cell product codebook limited feedback scheme,
if per-user feedback-bits (i.e., $B_{k}$) does not scale with system SNR $\rho$, then for
sufficiently large $P_{max}$, the per-user throughput $R_{k}^{LF}$
tends to a constant and scales according to,
\begin{IEEEeqnarray}{rCl} \label{eqn:R_IFL}
R_{k}^{LF} =
\mathcal{O}\left(\frac{n_{R}B_{k}\ln2}{\left(Nn_{T}-n_{R}\right)Nn_{T}}\right),
~\forall k=1, 2, \cdots, K.
\end{IEEEeqnarray}
\end{Cor}
\begin{proof}
Please refer to Appendix
\ref{app:rate_const_IFL} for the proof.
\end{proof}

\begin{Rem} [Effects of Heterogeneous Path Losses] \textcolor{black}{Note that, the results given in {\em Theorem \ref{thm:rate_loss}} and {\em Corollary \ref{thm:Bk_scale}} above are similar to those results stated in Theorem 1 and Theorem 2 of reference \cite{Ravindran2008}. The major difference is the path loss effect term $g_{k}^{sum}$, which results from the different path losses from the $N$  cooperating BSs to the $k^{th}$ MS.}
\end{Rem}

\begin{Rem} [Scaling Laws for the Proposed Limited Feedback Design]
In the noise-limited regime, the minimum number of feedback-bits $B_{k}$ required to maintain a bounded per-user throughput loss, shall scale w.r.t. the number of cooperating BSs according to,
\begin{IEEEeqnarray}{rCl} \label{eqn:Bk_scaling_law}
B_{k} = \mathcal{O}\bigg( Nn_{T}n_{R}\log_{2}\big(\rho g_{k}^{sum} \big)\bigg).
\end{IEEEeqnarray}

Moreover, the residual CCI term in equation \eqref{eqn:RkLF} is negligible for the noise-limited case. Following the proof of {\em Theorem \ref{thm:rate_loss}}, we can show that in the noise-limited regime, the achievable per-user throughput of the proposed limited feedback scheme scales as,
\begin{IEEEeqnarray}{rCl} \label{eqn:Rk_noise}
R_{k}^{LF} =  \mathcal{O} \left( n_{R} \log_{2} \bigg( \frac{n_{R}\rho g_{k}^{sum}}{Nn_{T}}  \bigg) \right).
\end{IEEEeqnarray}

On the other hand, in the interference-limited regime, the achievable per-user throughput of the proposed per-cell product codebook based limited feedback scheme shall scale as,
\begin{IEEEeqnarray}{rCl} \label{eqn:R_IFL_N}
R_{k}^{LF} = \mathcal{O}\left( \frac{n_{R}B_{k}}{(Nn_{T})^{2}} \right).
\end{IEEEeqnarray}
\end{Rem}

\section{Numerical Results and Discussions}\label{sec:num_res_disc}
In this section, we shall study the performance of the proposed
per-cell product codebook limited feedback scheme and verify the analytical
results via simulations. We shall first compare the proposed
per-cell product codebook limited feedback scheme with several baseline schemes. {\em Baseline 1: joint-cell codebook approach}; {\em Baseline 2 and 3}: Givens rotation approach with different number of feedback-bits. In the Givens rotation approach, the two Givens parameters of
a Givens matrix are quantized with a two-dimensional vector
quantizer \cite{Roh2004, Sadrabadi2006, Long2008}.
Then, we proceed to study
the performance-complexity tradeoff of the proposed low-complexity
feedback indices selection algorithm, as well as the analytical results in Section
\ref{sec:perf_analy}. In the simulations, two-dimensional hexagonal
cellular model is considered, with a cell-radius of \textcolor{black}{$300$ meters and the carrier frequency is set to be $2$ GHz.} The path
loss model specified in \cite{Std:4G-EMD:16m} is used, i.e., $PL(dB)
= 130.19 + 37.6\log_{10}(d(km))$, with $8$ dB lognormal shadowing effects. Users are assumed to be uniformly distributed within the cooperating cells. We use \emph{interference-free SNR} to represent the receiving SNR at the cell edge of a single-cell single-MS scenario.

\subsection{Performance of the Proposed Limited Feedback Design} \label{subsec:baseline}
\begin{figure}
\centering
\includegraphics[scale=0.6]{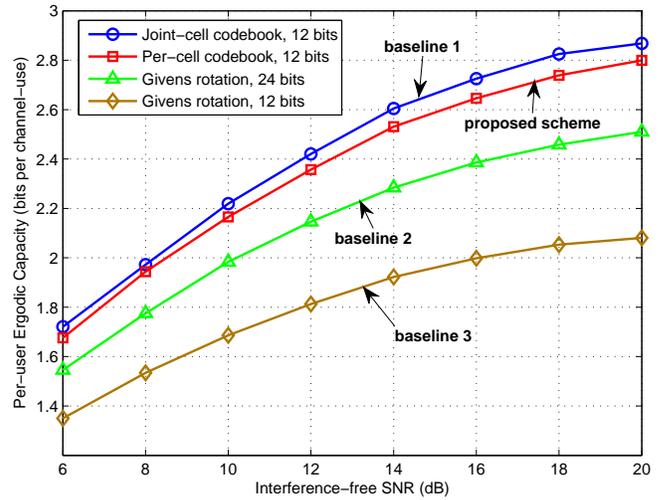}
\caption {Performance comparison of the proposed per-cell product codebook based limited feedback design with
joint-cell codebook and Givens rotation approaches, in a network MIMO system
$\left(n_{T},
 N, n_{R}, K\right)=\left(4, 3, 2, 6 \right)$. In the proposed scheme,
baseline 1 and baseline 3, we adopt 4 bits per BS for CSI feedback; while in
baseline 2, we adopt 8 bits per BS for CSI feedback. While the proposed
per-cell product codebook limited feedback scheme can achieve $95\%\sim97\%$ of
the performance of joint-cell codebook approach, it performs much better
than the Givens rotation approach.} \label{fig:per_comp}
\end{figure}

Fig. \ref{fig:per_comp} illustrates the per-user average throughput versus interference-free SNR. The simulation results show that, in the practical settings, the proposed per-cell
product codebook based limited feedback scheme can achieve $95\%\sim97\%$ the
performance of the joint-cell codebook approach. Moreover, the proposed scheme achieves much better throughput compared with {\em Baseline 2} (Givens rotation approach with doubled number of bits for limited feedback) and {\em Baseline 3} (Givens rotation approach). This is because the Givens rotation approach has quite low feedback efficiency due to two-dimensional vector quantization, compared with matrix quantization in
the codebook-based approach.

\subsection{Performance-complexity Tradeoff of the Low-complexity ISA} \label{subsec:comp_perfor}
In this section, we study the performance-complexity tradeoff of the
low-complexity ISA. Here, {\em performance} is measured in terms of
{\em per-user ergodic capacity}, and {\em complexity} is measured
in terms of {\em the number of codewords} that are searched for
generating the codewords indices. As stated in section
\ref{sec:local_search}, the parameters $\delta_{n}~(n=1,2,\cdots, N)$ determine the tradeoff between
performance and complexity of \emph{algorithm \ref{alg:
low_complex}}.
\begin{figure}
\centering
\includegraphics[scale = 0.6]{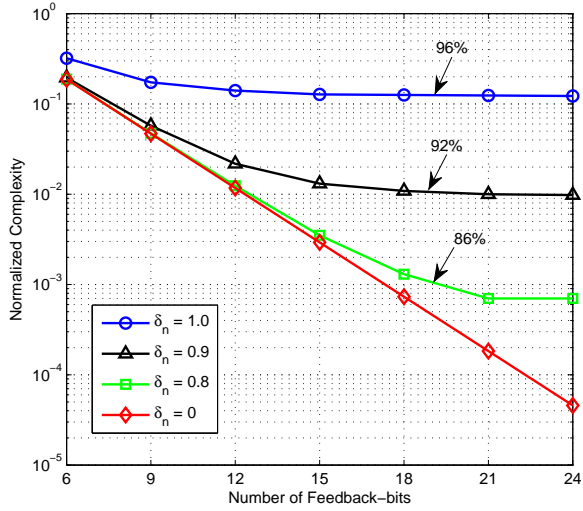}
\caption {Performance-complexity tradeoff of the proposed low-complexity feedback indices selection algorithm (\emph{algorithm
\ref{alg: low_complex}}), with different choices of $\delta_{n}$ for a
network MIMO system $\left(n_{T},
 N, n_{R}, K\right)=\left(4, 3, 2, 6 \right)$.  \textcolor{black}{The complexity numbers shown in the figure are relative complex defined in equation \eqref{eqn:R_Complx}.} Setting $\delta_{n} = 1, 0.9, 0.8$ can achieve $96\%$, $92\%$, $86\%$ the performance of exhaustive search,  with  $10\%$, $1\%$, $0.1\%$ the complexity of exhaustive search, respectively.
 Larger $\delta_{n}$ gives better performance at the cost of higher complexity, and vice versa. The curve with $\delta_{n}=0$ is  for reference.}
 \label{fig:cplx}
\end{figure}
Fig. \ref{fig:cplx} shows the performance-complexity tradeoff with
different choices of $\delta_{n}$, where we set $\delta_{n}$ to the
same value for all $n$. \textcolor{black}{The complexity numbers shown in the figure denote the relative complexity with respect to the exhaustive search method with original codebooks, i.e.,
\begin{IEEEeqnarray}{rCl} \label{eqn:R_Complx}
&&\mbox{Relative complexity}  \notag \\
&&~~~~\triangleq
\frac{ \mbox{No. of codewords searched with {\em Algorithm \ref{alg: low_complex}} } } {\mbox{No. of codewords searched with exhaustive search}} \notag \\
&&~~~~= \frac{\prod_{n=1}^{N}\left|\bar{\varphi}\left(\mathbf{V}_{k,n}^{(w)}, \delta_{n}\right)\right|}{ 2^{B_{k}} }.
\end{IEEEeqnarray}}
As shown in the figure, with $\delta_{n} = 0.9$, the low-complexity ISA can achieve
$92\%$ the performance of the exhaustive search approach, with only $1\%$
complexity of the exhaustive search. With larger $\delta_{n}$,
\emph{algorithm \ref{alg: low_complex}} can achieve better
performance at the cost of higher computational complexity, and vice versa.

\subsection{Verification of the Analytical Expressions} \label{subsec:scale_fb}
In this section, we shall compare the analytical results stated in
\emph{Corollary \ref{thm:Bk_scale}} and \emph{Corollary \ref{thm:rate_const_IFL}} with numerical results, and demonstrate the validity of those
analytical studies.
\begin{figure}
\centering
\includegraphics[scale=0.6]{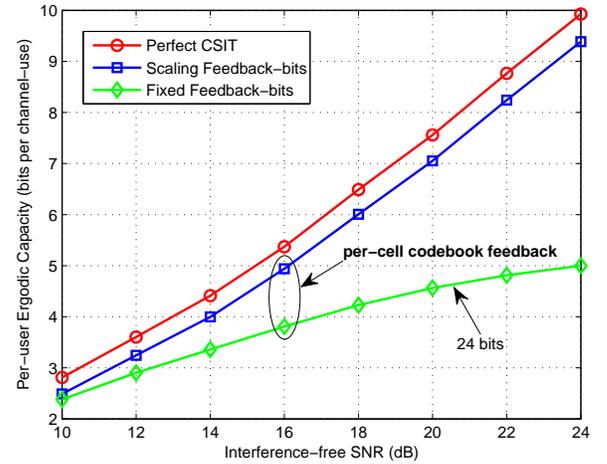}
\caption {\textcolor{black}{Per-user ergodic capacity versus interference-free SNR,
for perfect CSIT case, per-cell product codebook feedback with scaling
feedback-bits and fixed feedback-bits, with a system configuration
$\left(n_{T},
 N, n_{R}, K\right)=\left(8, 3, 2, 12 \right)$. For the scaling feedback-bits case,
 the per-user feedback-bits scale according to equation \eqref{eqn:Bk_approx}, and we set $\varepsilon = 1$. Specifically, the feedback-bits used in the scaling feedback-bits case are $[24  \:\ 30  \:\ 36 \:\  42 \:\  48 \:\   57 \:\  63  \:\ 72]$, corresponding to the $8$ SNR values respectively. A constant gap of about 0.5 (bits per channel use)
 between the scaling feedback and the perfect CSIT case is observed. For the fixed feedback-bits case, the system is interference-limited.}}
 \label{fig:scale_fb}
\end{figure}
Fig. \ref{fig:scale_fb} shows the simulation results for GCSI case, per-cell product codebook limited feedback with scaling
feedback-bits and fixed feedback-bits. For the per-cell product codebook based limited
feedback with scaling feedback-bits, we set $\varepsilon = 1$ and let the
feedback-bits scale according to equation \eqref{eqn:Bk_approx}.
Compared with GCSI case, about 0.5 (bits per channel use)
throughput loss is achieved with scaling-feedback.
%The reason why we
%get a gap of 0.5 rather than 1 is that equation
%\eqref{eqn:Bk_approx} is conservative, for the asymptotic throughput loss stated in \emph{Theorem
%\ref{thm:rate_loss}} is pessimistic.

\begin{figure}
\centering
\includegraphics[scale=0.6]{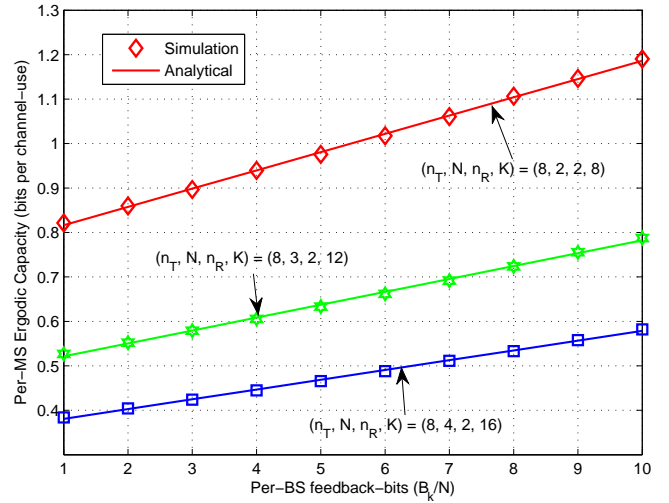}
\caption {Per-user throughput in the interference-limited regime
with different system configurations $\left(n_{T},
 N, n_{R}, K\right)$. The y-axis is the per-user ergodic capacity (in bits per channel use), and the x-axis is the per-BS feedback-bits.
 It is observed that the per-user throughput scale linearly with the feedback-bits, as stated in \emph{Corollary
\ref{thm:rate_const_IFL}}. The simulation results match
the analytical results stated in \emph{Corollary
\ref{thm:rate_const_IFL}}. } \label{fig:ifl_constant}
\end{figure}
Fig. \ref{fig:ifl_constant} shows the performance of the proposed
per-cell product codebook based limited feedback scheme in interference limited regime, with different number of
feedback-bits per BS (i.e., $\frac{B_{k}}{N}$). We can see that the per-user ergodic capacity scales linearly w.r.t. feedback-bits per BS  under all system configurations $\left(n_{T},
 N, n_{R}, K\right)$. The simulation results shown in Fig. \ref{fig:ifl_constant} match the
analytical results stated in \emph{Corollary
\ref{thm:rate_const_IFL}}.

%Fig. \ref{fig:macro_div} shows the simulation results of the per-user outage probability, with a target rate $R_{k}^{Tgt}$ of 2 bits per channel use. In the simulations, we set the interference-free SNR to be 18 dB. The figure shows that the per-user outage probability decays exponentially w.r.t. the number of cooperating BSs in noise limited regime, as stated in \emph{Theorem \ref{thm:outage_scale}}. The simulation results shown in Fig. \ref{fig:macro_div} match the
%analytical results stated in \emph{Theorem
%\ref{thm:outage_scale}} pretty well.

\section{Conclusions}\label{sec:conclusion}
In this paper, we have proposed a scalable per-cell product codebook based
limited feedback framework for network MIMO systems, along with a
low-complexity feedback indices selection algorithm. We have shown that the
proposed limited feedback design can asymptotically achieve the same performance as
the joint-cell codebook approach. When the number of per-user feedback-bits scales as $\mathcal{O}\big( Nn_{T}n_{R}\log_{2}(\rho g_{k}^{sum})\big)$, the proposed scheme operates in a noise-limited regime with a per-user throughput scaling as $\mathcal{O}\left( n_{R} \log_{2} \big( \frac{n_{R}\rho g_{k}^{sum}}{Nn_{T}}  \big) \right)$. On the other hand, when the number of per-user feedback-bits does not scale with system SNR, the system operates in an interference-limited regime with a per-user throughput scaling as $\mathcal{O}\left( \frac{n_{R}B_{k}}{(Nn_{T})^{2}} \right)$.
%In addition, we have shown that the per-user outage probability scales as $\mathcal{O}\left( \big( %\frac{n_{R}\rho g_{k}}{n_{T}} \big )^{-N} \right)$ in the noise limited regime.
The
numerical results show that the proposed scheme can achieve similar
performance as the joint-cell codebook approach and performs much
better than the Givens rotation approach in practical settings. One
interesting direction for further study is to consider adaptive
feedback-bits allocation based on users path loss geometry.

\appendices
\section{Proof of Lemma \ref{lem: add_const}} \label{app: L2_add_const}
\textcolor{black}{For ease of elaboration, we first introduce the following intermediate lemma.}
\begin{Lem} \label{lem:vector_as}
\textcolor{black}{Consider a random vector  $\mathbf{h} \in \mathbb{C}^{Nn_{T}}$ with i.i.d. $\mathbb{CN}(0, \sigma^{2})$ entries, and let
$\mathbf{\tilde{h}}=\frac{\mathbf{h}}{\|\mathbf{h}\|}$ denote the
direction of $\mathbf{h}$. Partitioning $\mathbf{\tilde{h}}$ into
$N$ \emph{sub-vectors}, i.e.,
$\mathbf{\tilde{h}}=\left[\mathbf{\tilde{h}}_{1}^{\dag} \:\
\mathbf{\tilde{h}}_{2}^{\dag} \:\ \cdots \:\
\mathbf{\tilde{h}}_{N}^{\dag} \right]^{\dag}$, where
$\mathbf{\tilde{h}}_{n} \in \mathbb{C}^{n_{T}}$,  $\forall n =1, 2,
\cdots, N$, we then have,
\begin{IEEEeqnarray}{rCl} \label{eqn:vector_as}
\mbox{Pr}\left\{\lim_{n_{T}\rightarrow
\infty}\left\{\mathbf{\tilde{h}}_{n}^{\dag}\mathbf{\tilde{h}}_{n}
\right\}= \frac{1}{N}\right\}=1, ~\forall~n=1,2,\cdots, N.
\end{IEEEeqnarray}}
\end{Lem}
{\em Proof}: \textcolor{black}{
Denote $\tilde{h}_{i}$ as the $i^{th}$ element of
$\mathbf{\tilde{h}}$. Since $\mathbf{h}$ is a random vector with
i.i.d. $\mathbb{CN}(0, \sigma^{2})$ entries, all $\tilde{h}_{i}~(i
=1 ,2, \cdots, Nn_{T})$ are identically distributed and satisfy,
\begin{IEEEeqnarray}{rCl}
\mbox{\emph{Var}}\left\{\tilde{h}_{i}\tilde{h}_{i}^{\dag}\right\}=\frac{Nn_{T}-1}{\left(Nn_{T}\right)^2\left(Nn_{T}+1\right)},~\forall
~ i = 1, 2, \cdots, Nn_{T},
\end{IEEEeqnarray}
where
$\mbox{\emph{Var}}\left\{\tilde{h}_{i}\tilde{h}_{i}^{\dag}\right\}$
denotes the variance of $\tilde{h}_{i}\tilde{h}_{i}^{\dag}$.
Moreover, it is straightforward to get that
$\mathbb{E}\left\{\mathbf{\tilde{h}}_{n}^{\dag}\mathbf{\tilde{h}}_{n}\right\}=\frac{1}{N}$,
and $\mbox{\emph{Var}}\left\{\mathbf{\tilde{h}}_{n}^{\dag}\mathbf{\tilde{h}}_{n}\right\}=\frac{N-1}{N^2\left(Nn_{T}+1\right)},
~\forall~n=1, 2, \cdots, N$.}

\textcolor{black}{Define
$\psi_{t}=\left\{\left|\mathbf{\tilde{h}}_{n}^{\dag}\mathbf{\tilde{h}}_{n}-\frac{1}{N}\right|>\epsilon\right\}$,
where $\epsilon$ denotes any given positive number. Using Chebyshev's inequality, we get,
\begin{IEEEeqnarray}{rCl}
\mbox{Pr}\left\{\psi_{t}\right\} =
\mbox{Pr}\left\{\left|\mathbf{\tilde{h}}_{n}^{\dag}\mathbf{\tilde{h}}_{n}-\frac{1}{N}\right|>\epsilon\right\}
\leq \frac{N-1}{N^2\left(Nt+1\right)\epsilon^2},
\end{IEEEeqnarray}
which implies that \cite[page 37]{Sen1993},
\begin{IEEEeqnarray}{rCl}
\mbox{Pr}\left\{\bigcup_{t \geq n_{T}}\psi_{t}\right\} \leq \sum_{t
\geq n_{T}} \mbox{Pr}\left\{\psi_{t}\right\}  \rightarrow 0
~\mbox{ as }~ n_{T} \rightarrow \infty,
\end{IEEEeqnarray}
which proves {\em Lemma \ref{lem:vector_as}} \cite[pape 34-35]{Sen1993}.} ~\hfill\IEEEQEDclosed
%\end{proof}

\textcolor{black}{We next proceed to prove the first statement of {\em Lemma \ref{lem: add_const}}. We first partition $\mathbf{V}_{k}^{(w)}\in
\mathbb{C}^{Nn_{T}\times n_{R}}$ into $N$ sub-matrices, i.e.,
\begin{IEEEeqnarray}{rCl} \label{eqn:sub_matrices}
\mathbf{V}_{k}^{(w)} = \left[
\left[\mathbf{V}_{k,1}^{(w)}\right]^{\dag} \:
\left[\mathbf{V}_{k,2}^{(w)}\right]^{\dag} \: \cdots \:
\left[\mathbf{V}_{k,N}^{(w)}\right]^{\dag} \right]^{\dag},
\end{IEEEeqnarray}
where $\mathbf{V}_{k,n}^{(w)} \in \mathbb{C}^{n_{T} \times n_{R}},~
\forall ~n=1,2,\cdots, N$. Then, as a direct consequence of  \emph{Lemma \ref{lem:vector_as}},
\begin{IEEEeqnarray}{rCl} \label{eqn:a_s}
\mbox{Pr}\left\{\lim_{n_{T}\rightarrow
\infty}\left\{\left[\mathbf{V}_{k,n}^{(w)}\right]^{\dag}\mathbf{V}_{k,n}^{(w)}
\right\}= \frac{1}{N}\mathbf{I}_{n_{R}}\right\}=1,
\end{IEEEeqnarray}}

\textcolor{black}{Equation \eqref{eqn:a_s} suggests that when $n_{T}$ is sufficiently large, the orthonormal basis $\mathbf{V}_{k}^{(w)}$
shall have the same structure as the \emph{aggregate-codeword} defined in \eqref{eqn:aggregate} almost surely (i.e., with probability $1$), which proves the first statement of \emph{Lemma \ref{lem:vector_as}}.}

\textcolor{black}{Note that, the first statement of \emph{Lemma \ref{lem: add_const}} in turn suggests that the codewords in the joint-cell codebook shall have the same structure as the \emph{aggregate-codeword} defined in \eqref{eqn:aggregate} almost surely for sufficiently large $n_{T}$ and finite $N$. As a result, the second statement of \emph{Lemma \ref{lem: add_const}} can be derived from the definition of average distortion associated with joint-cell codebooks and per-cell product codebooks given in equation \eqref{eqn:ave_dist_joint} and \eqref{eqn:ave_dist_per}, respectively. Therefore, the expected distortion (i.e., average distortion) associated with the random per-cell product codebooks shall be the same as the expected distortion (i.e., average distortion) with random joint-cell codebooks.}

\section{Proof of Lemma \ref{lem:approx_error}}\label{app:approx_error}
By virtue of \emph{Lemma \ref{lem: add_const}}, for large $B_{k}$, the average quantization distortion associated with the random per-cell product codebooks can be approximated as \cite{Dai2008},
\begin{IEEEeqnarray}{rCl} \label{eqn:dist_upbd}
\bar{D}_{k}\left( \Phi_{per} \right ) \approx
\frac{\Gamma\left(\frac{1}{\alpha}\right)}{\alpha}\beta^{-\frac{1}{\alpha}}2^{-\frac{B_{k}}{\alpha}}+o(1)
\end{IEEEeqnarray}
where $\alpha=n_{R}\left(Nn_{T}-n_{R}\right)$;
$\beta=\frac{1}{\alpha!}\prod_{i=1}^{n_{R}}\frac{\left(Nn_{T}-i\right)!}{\left(n_{R}-i\right)!}$;
$\Gamma\left(x\right)$ denotes the Gamma function; the $o(1)$ term
can be ignored when $B_{k}$ is large or $n_{R}$ is small
\cite{Dai2008}. Since
$\lim_{\alpha\rightarrow\infty}\frac{\Gamma\left(\frac{1}{\alpha}\right)}{\alpha}=1$,
for large $\alpha$ (which is true in network MIMO system), we have $
\frac{\Gamma\left(\frac{1}{\alpha}\right)}{\alpha}\approx 1$.

Substituting Stirling's approximation for factorial, we get $\beta
\approx  \frac{\left(Nn_{T}\right)^{\frac{n_{R}-1}{2}}
}{\left(n_{R}\right)^{Nn_{R}n_{T}}}$ and,
\begin{IEEEeqnarray}{rCl} \label{eqn:beta_exp}
\beta^{-\frac{1}{\alpha}} \approx
\frac{n_{R}}{\left(Nn_{T}\right)^{\frac{n_{R}-1}{2Nn_{R}n_{T}}}}
\stackrel{(a)}{\approx} n_{R},
\end{IEEEeqnarray}
where (a) is because of that
$\lim_{n_{T}\rightarrow\infty}\left(Nn_{T}\right)^{\frac{n_{R}-1}{2Nn_{R}n_{T}}}=1$
and $Nn_{T}$ is usually very large for network MIMO system. Therefore, the approximation of the average quantization distortion associated with the random per-cell product codebooks $\bar{D}_{k}\left( \Phi_{per} \right )$ can be further simplified as $\bar{D}_{k}\left( \Phi_{per} \right ) \approx
n_{R}2^{-\frac{B_{k}}{n_R\left(Nn_{T}-n_{R}\right)}}$.

\section{Proof of Theorem \ref{thm:rate_loss}} \label{app: T1_rate_loss}
Here is the proof of {\em Theorem \ref{thm:rate_loss}}.
\begin{IEEEeqnarray}{rCl} \label{eqn:t1_proof}
R_{k}^{Loss} &\stackrel{(a)}{\leq}& \mathbb{E}\big\{\log_{2}\det\big(\mathbf{I}_{n_{R}} \notag \\
&+&\frac{1}{\sigma^{2}}\mathbf{H}_{k}\sum_{j=1,j\neq
k}^{K}\mathbf{\widehat{W}}_{j}\mathbf{P}_{j}\mathbf{\widehat{W}}_{j}^{\dag}\mathbf{H}_{k}^{\dag}\big)\big\}
\notag \\
&\stackrel{(b)}{=}&\mathbb{E}\bigg\{\log_{2}\det\big(\mathbf{I}_{n_{R}}+\frac{1}{\sigma^{2}}
\big[\mathbf{V}_{k}^{(w)}
\big]^{\dag}\mathbf{G}_{k} \notag \\
&&\sum_{j=1,j\neq
k}^{K}\mathbf{\widehat{W}}_{j}\mathbf{P}_{j}\mathbf{\widehat{W}}_{j}^{\dag}\mathbf{G}_{k}\mathbf{V}_{k}^{(w)}\left[\mathbf{S}_{k}^{(w)}\right]^2
\big)\bigg\} \notag \\
&\stackrel{(c)}{\leq}&\log_{2}\det\big(\mathbf{I}_{n_{R}}+\frac{1}{\sigma^{2}}\mathbb{E}\bigg\{
\big[\mathbf{V}_{k}^{(w)}
\big]^{\dag}\mathbf{G}_{k} \notag \\
&&\sum_{j=1,j\neq
k}^{K}\mathbf{\widehat{W}}_{j}\mathbf{P}_{j}\mathbf{\widehat{W}}_{j}^{\dag}\mathbf{G}_{k}\mathbf{V}_{k}^{(w)}\left[\mathbf{S}_{k}^{(w)}\right]^2
\bigg\}\big), \notag
\end{IEEEeqnarray}
where (a) and (c) are obtained following the approaches in \cite{Ravindran2008}; (b) follows by substituting equation
\eqref{eqn:svd_Hw} for $\mathbf{H}_{k}^{(w)}$.

Let $\mathbf{F}_{k}=\mathbb{E}\bigg\{ \bigg[\mathbf{V}_{k}^{(w)}
\bigg]^{\dag}\mathbf{G}_{k}\sum_{j=1,j\neq
k}^{K}\mathbf{\widehat{W}}_{j}\mathbf{P}_{j}\mathbf{\widehat{W}}_{j}^{\dag}\mathbf{G}_{k}\mathbf{V}_{k}^{(w)}$\\$\bigg[\mathbf{S}_{k}^{(w)}\bigg]^2
\bigg\}$, and note that $\mathbf{V}_{k}^{(w)}$, $\mathbf{G}_{k}$,
$\mathbf{\widehat{W}}_{j}$ and $\left[\mathbf{S}_{k}^{(w)}\right]^2$
are mutually independent, the expectation can be carried out step by
step.
\begin{itemize}
\item Step 1: substituting the decomposition of
$\mathbf{V}_{k}^{(w)}$, i.e., $\mathbf{V}_{k}^{(w)} =
\mathbf{\widehat{V}}_{k}^{(w)}\mathbf{X}_{k}\mathbf{Y}_{k}+\mathbf{\widetilde{V}}_{k}^{(w)}\mathbf{Z}_{k}$~(see
\emph{Lemma 1} of \cite{Ravindran2008}).
\begin{IEEEeqnarray}{rCl} \label{eqn:Fk2}
\mathbf{F}_{k}=~&& \mathbb{E}\left\{ \left[\mathbf{V}_{k}^{(w)}
\right]^{\dag}\mathbf{G}_{k}\sum_{j=1,j\neq
k}^{K}\mathbf{\widehat{W}}_{j}\mathbf{P}_{j}\mathbf{\widehat{W}}_{j}^{\dag}\mathbf{G}_{k}\mathbf{V}_{k}^{(w)}\right\}\notag \\
~&&\mathbb{E}^{(1)}\left\{
\left[\mathbf{S}_{k}^{(w)}\right]^2 \right\} \notag \\
\stackrel{(d)}{=}~&& Nn_{T}\mathbb{E}\left\{
\mathbf{Z}_{k}^{\dag}\bigg[\mathbf{\widetilde{V}}_{k}^{(w)}\right]^{\dag}\mathbf{G}_{k} \notag \\
~&&\sum_{j=1,j\neq
k}^{K}\mathbf{\widehat{W}}_{j}\mathbf{P}_{j}\mathbf{\widehat{W}}_{j}^{\dag}\mathbf{G}_{k}\mathbf{\widetilde{V}}_{k}^{(w)}\mathbf{Z}_{k}\bigg\},
\end{IEEEeqnarray}
where (d) is because of that $\mathbb{E}^{(1)}\left\{
\left[\mathbf{S}_{k}^{(w)}\right]^2
\right\}=Nn_{T}\mathbf{I}_{n_{R}}$~(see Appendix B of
\cite{Ravindran2008}) and we have used the LF-BD conditions
\eqref{eqn:bd}.

\item Step 2: calculating the expectation of $\left(\sum_{j=1,j\neq
k}^{K}\mathbf{\widehat{W}}_{j}\mathbf{P}_{j}\mathbf{\widehat{W}}_{j}^{\dag}\right)$. Let
\begin{IEEEeqnarray}{rCl}
\mathbf{Q}_{k}= \mathbb{E}^{(2)}\left\{\sum_{j=1,j\neq
k}^{K}\mathbf{\widehat{W}}_{j}\mathbf{P}_{j}\mathbf{\widehat{W}}_{j}^{\dag}\right\},
\end{IEEEeqnarray}
where $\mathbb{E}^{(2)}$ denotes expectation taken over the
distribution of $\mathbf{\widehat{W}}_{j}$. When the number of
active users is large and the users are randomly distributed, we can
safely conclude that $\mathbf{Q}_{k} \approx
\frac{n_{R}\left(K-1\right)}{Nn_{T}}p\mathbf{I}_{Nn_{T}}$, with $p = \frac{NP_{max}}{Kn_{R}}$.

\item Step 3: calculating expectation over $\mathbf{\widetilde{V}}_{k}^{(w)}$.
\begin{IEEEeqnarray}{rCl}
\mathbb{E}^{(3)}\left\{\left[\mathbf{\widetilde{V}}_{k}^{(w)}
\right]^{\dag}\mathbf{G}_{k}\mathbf{Q}_{k}\mathbf{G}_{k}\mathbf{\widetilde{V}}_{k}^{(w)}\right\}=\frac{pn_{R}\gamma_{k}\left(K-1\right)}{\left(Nn_{T}\right)^{2}}\mathbf{I}_{n_{R}},
\end{IEEEeqnarray}
where $\mathbb{E}^{(3)}$ denotes expectation taken over the
distribution of $\mathbf{\widetilde{V}}_{k}^{(w)}$ (isotropic
distribution), and
$\gamma_{k}=n_{T}\mathop{\sum}_{n=1}^{N}g_{k,n}s_{k,n}$.

\item Step 4: calculating expectation over lognormal-shadowing. Let
\begin{IEEEeqnarray}{rCl}
\tilde{\gamma}_{k}=\mathbb{E}^{(4)}\left\{\gamma_{k}\right\}=n_{T}g_{k}^{sum},
\end{IEEEeqnarray}
where $\mathbb{E}^{(4)}$ denotes expectation taken over
lognormal-shadowing.

\item Step 5: calculating expectation over quantization error
$\mathbf{Z}_{k}$. We have,
\begin{IEEEeqnarray}{rCl} \label{eqn:beta}
\frac{n_{R}\tilde{\gamma}_{k}}{\left(Nn_{T}\right)^{2}}p\mathbb{E}^{(5)}\left\{
\mathbf{Z}_{k}^{\dag}\mathbf{Z}_{k} \right\}
\stackrel{(e)}{=}\frac{p\tilde{\gamma}_{k}\bar{D}\left(K-1\right)}{\left(Nn_{T}\right)^{2}}\mathbf{I}_{n_{R}},
\end{IEEEeqnarray}
where $\mathbb{E}^{(5)}$ denotes expectation taken over the
distribution of $\mathbf{Z}_{k}$, and (e) is given in Appendix B of
\cite{Ravindran2008} with $\bar{D} = n_{R}2^{-\frac{B_{k}}{n_R\left(Nn_{T}-n_{R}\right)}}$.

\item Step 6: Finally, substituting equation \eqref{eqn:beta} into \eqref{eqn:Fk2},
we get,
\begin{IEEEeqnarray}{rCl} \label{eqn:prf_RkLoss}
\mathbf{F}_{k} =
\frac{p\tilde{\gamma}_{k}}{Nn_{T}}\bar{D}\mathbf{I}_{n_{R}}=\frac{p\bar{D}\left(K-1\right)g_{k}^{sum}}{N}\mathbf{I}_{n_{R}}.
\end{IEEEeqnarray}
\end{itemize}

Therefore, the asymptotic per-user throughput loss due to limited feedback is given by,
\begin{IEEEeqnarray}{rCl} \label{eqn:rate_loss_final}
R_{k}^{Loss} = \mathcal{O} \left( n_{R}\log_{2} \big(  2^{-\frac{B_{k}}{n_{R}(Nn_{T}-n_{R})}} \rho g_{k}^{sum} \big) \right).
\end{IEEEeqnarray}

\section{Proof of Corollary \ref{thm:rate_const_IFL}} \label{app:rate_const_IFL}
The proof of  {\em Corollary \ref{thm:rate_const_IFL} } can be summarized as follows.
\begin{IEEEeqnarray}{rCl} \label{eqn:R_LF}
R_{k}^{IFL}
&\stackrel{(a)}{\approx}&\mathbb{E}\left\{\log_{2}\det\left(\mathbf{H}_{k}\sum_{j=1}^{K}\mathbf{\widehat{W}}_{j}\mathbf{\widehat{W}}_{j}^{\dag}\mathbf{H}_{k}^{\dag}\right)\right\}
\notag \\
~~& -&\mathbb{E}\left\{\log_{2}\det\left(\mathbf{H}_{k}\sum_{j=1,j\neq
k}^{K}\mathbf{\widehat{W}}_{j}\mathbf{\widehat{W}}_{j}^{\dag}\mathbf{H}_{k}^{\dag}\right)\right\}\notag\\
& \stackrel{(b)}{\approx}& \log_{2}\det\left(\mathbb{E}\left\{
\mathbf{H}_{k}\sum_{j=1}^{K}\mathbf{\widehat{W}}_{j}\mathbf{\widehat{W}}_{j}^{\dag}\mathbf{H}_{k}^{\dag}
\right\}\right) \notag \\
~~&-&\log_{2}\det\left(\mathbb{E}\left\{
\mathbf{H}_{k}\sum_{j=1,j\neq
k}^{K}\mathbf{\widehat{W}}_{j}\mathbf{\widehat{W}}_{j}^{\dag}\mathbf{H}_{k}^{\dag}
\right\} \right) \notag \\
&\stackrel{(c)}{=}&\mathcal{O}\left( n_{R}\log_{2}\left( 1 +
\frac{2^{\frac{B_{k}}{n_{R}\left(Nn_{T}-n_{R}\right)}}}{K-1} \right)
\right) \notag \\
&\stackrel{(d)}{=}&\mathcal{O}\left(\frac{B_{k}\ln2}{\left(Nn_{T}-n_{R}\right)\left(K-1\right)}\right),
\end{IEEEeqnarray}
where (a) is because the noise term is negligible compared with the
CCI term; (b) holds in the \emph{orderwise} sense; (c) follows the same arguments as in the proof of \emph{Theorem
\ref{thm:rate_loss}}; (d) is because
$\frac{B_{k}}{n_{R}\left(Nn_{T}-n_{R}\right)}$ is quite small and
$2^{\frac{B_{k}}{n_{R}\left(Nn_{T}-n_{R}\right)}}\approx1 +
\frac{B_{k}\ln2}{n_{R}\left(Nn_{T}-n_{R}\right)}$.

\bibliographystyle{IEEEtran}
% argument is your BibTeX string definitions and bibliography database(s)
\bibliography{IEEEabrv,references_cited}
\begin{IEEEbiography}
%[{\includegraphics[width=1in,height=1.05in,clip,keepaspectratio]{yong_cheng_photo.eps}}]
{Yong Cheng (S'09)}
received the B.Eng. (1st honors) from Zhejiang University (2002-2006), Hangzhou, China, and the MPhil in Electronic and Computer Engineering  from the Hong Kong
University of Science and Technology (HKUST) (2008-2010), Hong Kong.  His research interests
include limited feedback design for MIMO/MISO systems,  resource control and optimization in wireless networks, cooperative MIMO systems, as well as relay networks.
\end{IEEEbiography}
\begin{IEEEbiography}
%[{\includegraphics[width=1in,height=1.25in,clip,keepaspectratio]{vincent_lau_photo.eps}}]
{Vincent Lau (SM'04)}
obtained B.Eng (Distinction 1st Hons) from the University of Hong
Kong (1989-1992) and Ph.D. from Cambridge University (1995-1997). He
was with HK Telecom (PCCW) as system engineer from 1992-1995 and
Bell Labs - Lucent Technologies as member of technical staff from
1997-2003. He then joined the Department of ECE, Hong Kong
University of Science and Technology (HKUST) as Associate Professor.
His current research interests include the robust and
delay-sensitive cross-layer scheduling of MIMO/OFDM wireless systems
with imperfect channel state information, cooperative and cognitive
communications, dynamic spectrum access as well as stochastic
approximation and Markov Decision Process.
\end{IEEEbiography}
\begin{IEEEbiography}
%[{\includegraphics[width=1in,height=1.25in,clip,keepaspectratio]{yi_long_photo.eps}}]
{Yi Long} received his Ph.D. degree from Tsinghua University, Beijing, China in 2008. From 2008 to 2010 he worked on cellular cooperative communications. He is now an engineer at Huawei Technology Co. Ltd., Shenzhen, China. His research interests include interference management, MIMO, and iterative algorithms.
\end{IEEEbiography}
\end{document}